\documentclass[a4paper]{jpconf}
\usepackage{graphicx}
\usepackage{wrapfig}
\begin{document}
\title{Cross section and double-helicity asymmetry in charged hadron production in $p+p$ collisions at $\sqrt{s}=62.4$~GeV at PHENIX}

\author{Christine A Aidala, for the PHENIX Collaboration}

\address{Los Alamos National Laboratory, Los Alamos, NM 87545, U.S.A.}

\ead{caidala@bnl.gov}

\begin{abstract}
The cross section and double-helicity asymmetry for production of non-identified positive and negative charged hadrons at midrapidity in $p+p$ collisions at $\sqrt{s} = 62.4$~GeV have been measured by the PHENIX experiment at the Relativistic Heavy Ion Collider (RHIC) for a transverse momentum range of 0.5--4.5~GeV/$c$.  The cross section measurements are compared to next-to-leading order and next-to-leading log perturbative QCD calculations, providing information on the applicability of these calculation techniques in the measured kinematic range.  The double-helicity asymmetry measurement at this moderate energy provides sensitivity to $\Delta G$, the gluon spin contribution to the spin of the proton, up to modestly larger values of the gluon momentum fraction than previous RHIC measurements at 200~GeV.
\end{abstract}

\section{Introduction}

The production of hadrons at large transverse momentum offers a classical test of perturbative QCD (pQCD).  At fixed-target center-of-mass energies of $\sim20-40$~GeV, next-to-leading order (NLO) pQCD calculations underpredict experimental results by factors of 1.5 or more.  A technique called threshold resummation, which recovers a portion of the higher-order terms in the strong coupling constant expansion, has been demonstrated to improve agreement with lower-energy data, and the techniques for applying cuts in calculations to match experimental data have been recently improved (see~\cite{Almeida:2009jt} and the discussion and references therein).  It is interesting to test the relevance of this technique at energies between those of fixed-target experiments and typical collider energies.

While cross sections for midrapidity particle production in $\sqrt{s} = 200$~GeV proton-proton collisions have been described reasonably well by NLO pQCD calculations, a measurement by PHENIX of midrapidity neutral pion production in $p+p$ collisions at $\sqrt{s} = 62.4$~GeV suggests that next-to-leading log (NLL) resummation may improve the agreement of theory with data~\cite{Adare:2008qb}. Measurement of the cross section for production of non-identified charged hadrons from the same data set provides an independent measurement that is sensitive to different fragmentation functions (FFs) than the neutral pion measurement and allows further comparison of pQCD calculations with data at this intermediate center-of-mass energy.

\begin{figure}[h]
\begin{minipage}{18pc}
\includegraphics[width=18pc]{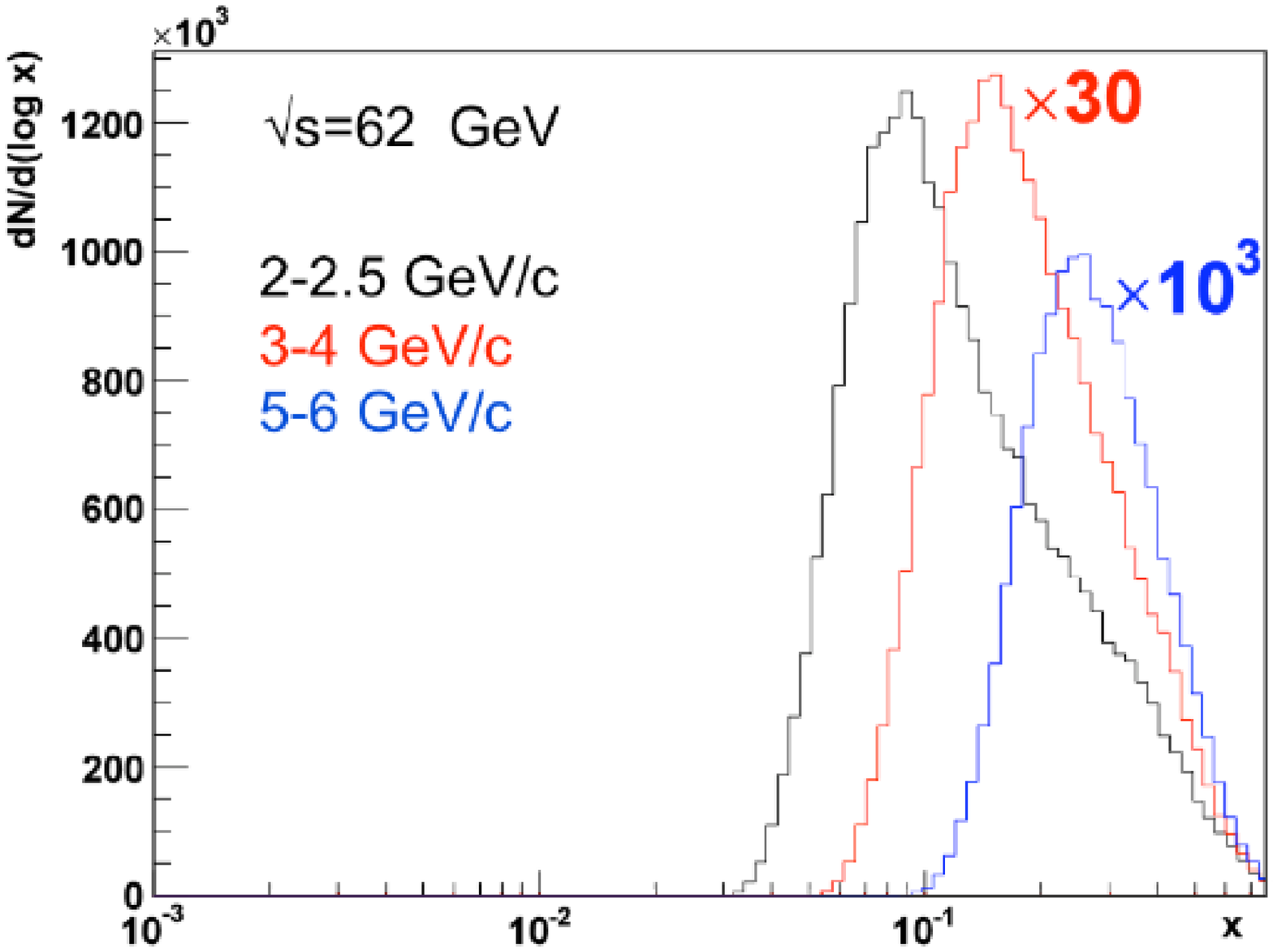}
\caption{\label{fig:xRange62GeV}$x$ range sampled by midrapidity pion production at $\sqrt{s}=62.4$~GeV for $p_T$ 2--2.5, 3--4, and 5--6~GeV/$c$.}
\end{minipage}\hspace{1.5pc}%
\begin{minipage}{18pc}
\vspace{1.5pc}
\includegraphics[width=17pc]{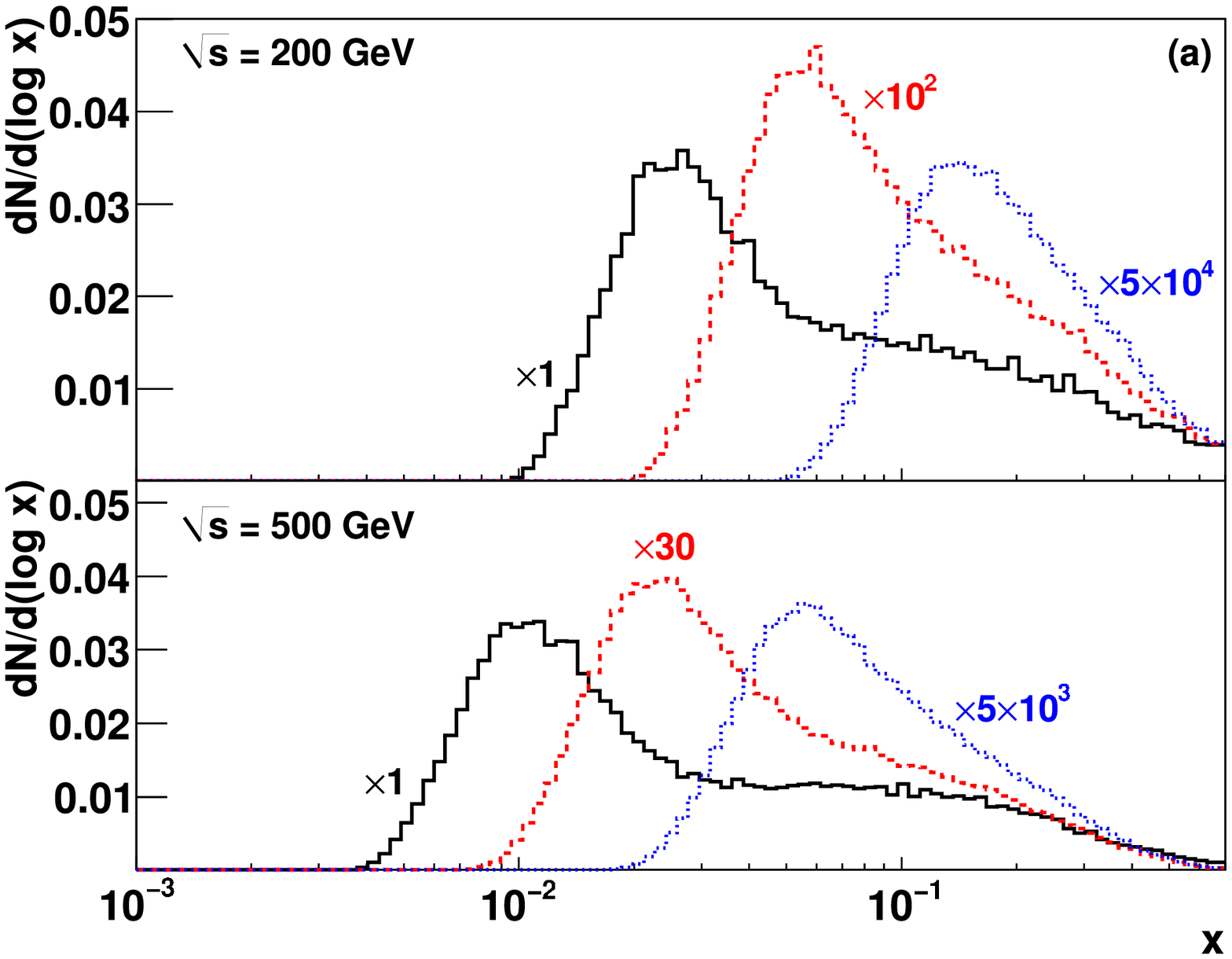}
\caption{\label{fig:xRange200_500GeV}$x$ range sampled by midrapidity pion production at $\sqrt{s}=200$~GeV (top) and 500~GeV (bottom) for $p_T$ 2--2.5, 4--5, and 9--12~GeV/$c$.}
\end{minipage}
\end{figure}

The cross section results also provide support for interpretation of the double-helicity asymmetry ($A_{LL}$) in charged hadron production, sensitive to the gluon spin contribution to the proton ($\Delta G$).  Compared to previous midrapidity $A_{LL}$ measurements from RHIC at $\sqrt{s} = 200$~GeV sensitive to $\Delta G$~\cite{Adare:2007dg,Abelev:2007vt,Adare:2008px,Abelev:2009pb,Adare:2010cy}, measuring at $\sqrt{s}=62.4$~GeV can access slightly higher values of the gluon momentum fraction, $x$, as can be seen by comparing Fig.~\ref{fig:xRange62GeV} and the top panel of Fig.~\ref{fig:xRange200_500GeV}.  Future RHIC measurements at the higher center-of-mass energy of 500~GeV are expected in turn to allow sensitivity to smaller gluon $x$ values, as shown in the bottom panel of Fig.~\ref{fig:xRange200_500GeV}.

\section{Data and analysis}

The data were taken by the PHENIX experiment at RHIC in 2006.  15.6~nb$^{-1}$ were used for the cross section analysis, 11.9~nb$^{-1}$ of which were longitudinally polarized data used for the $A_{LL}$ analysis.  Coincident hits in two Beam-Beam Counters (BBC) provided a minimum-bias (MB) trigger as well as the event vertex position.  Tracks were measured in the two Central Arms, each covering $\Delta \phi = \pi$ and $|y| < 0.35$.  The Drift Chamber (DC) and Pad Chambers (PC) were used to reconstruct the charged tracks.  A veto on the Ring-Imaging Cherenkov Detector (RICH) was required in order to eliminate background from electrons.  For a full description of the PHENIX detector see~\cite{Adcox:2003zm}.

\begin{figure}
\begin{center}
\includegraphics[width=0.7\textwidth]{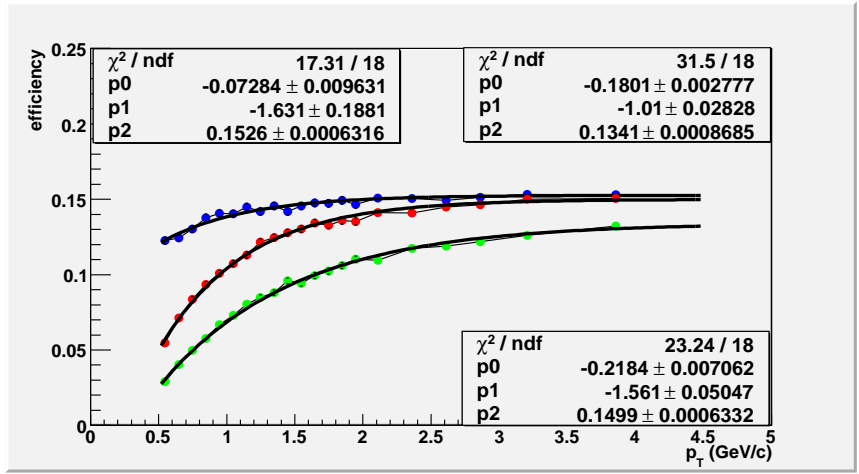}
\caption{\label{fig:eff}Absolute efficiencies for negative pions (top curve), negative kaons (bottom curve), and antiprotons (middle curve) estimated from single-particle Monte Carlo simulations generated over 2$\pi$ in azimuth and one unit of rapidity.}
\end{center}
\end{figure}

Tracking in the 2006 PHENIX detector configuration only started at a radial distance of 2~m from the interaction point, the inner radius of the DC.  As such the detection efficiency for pions and kaons was reduced due to decays in flight.  The absolute detection efficiencies for each particle species were estimated via Monte Carlo simulations.  Single particles were generated over $2\pi$ in azimuth and one unit in rapidity, with a vertex position distribution that reproduced that of the real data, then passed through a full GEANT 3.21~\cite{GEANT} detector simulation.  The efficiencies for negative particles are shown in Fig.~\ref{fig:eff}; those for positive particles were similar.  The efficiencies for the particle species increase as a function of $p_T$ and differ from one another due to decays in flight in the case of the pions and kaons, as well as due to mass effects, since all species were generated with a flat distribution over one unit of rapidity.

In order to apply the species-dependent efficiency corrections as a function of $p_T$, the produced species ratios from other measurements were used as input.  The species ratios were obtained from simultaneous fits to identified charged hadron ratios at midrapidity from the same PHENIX data set at $\sqrt{s}=62.4$~GeV~\cite{Adare:2010np} as well as from the British-Scandinavian Collaboration at the CERN ISR at $\sqrt{s}=52.8$ and 63.0~GeV~\cite{Alper:1974rw}.  $x_T$ scaling was used to approximately correct the ISR measurements to $\sqrt{s}=62.4$~GeV.

Background contributions came primarily from long-lived particle decays and were estimated to be 5\% or less for $p_T <2.75$~GeV/$c$, rising to $\sim 10$\% for $3.25 < p_T < 3.5$~GeV/$c$ and $\sim 30$\% for $4.0 < p_T < 4.5$~GeV/$c$.  For the cross section analysis, these background contributions were subtracted statistically.  For the asymmetry analysis, the asymmetry for tracks that hit the outer PC, located a radial distance of 5~m from the interaction point, between 3 and 9 sigma in $\phi$ from the expected hit position based on a projection from the DC.  The asymmetry of these background tracks was then subtracted from that calculated for the tracks within 2 sigma of the expected hit position in $\phi$ and $z$, weighted by the background fraction in that $p_T$ bin.  No corrections to the cross section or asymmetry were performed for feed-down from hyperon decays.

\section{Results and discussion}

\begin{figure}[hb]
\begin{minipage}{18pc}
\includegraphics[width=18pc]{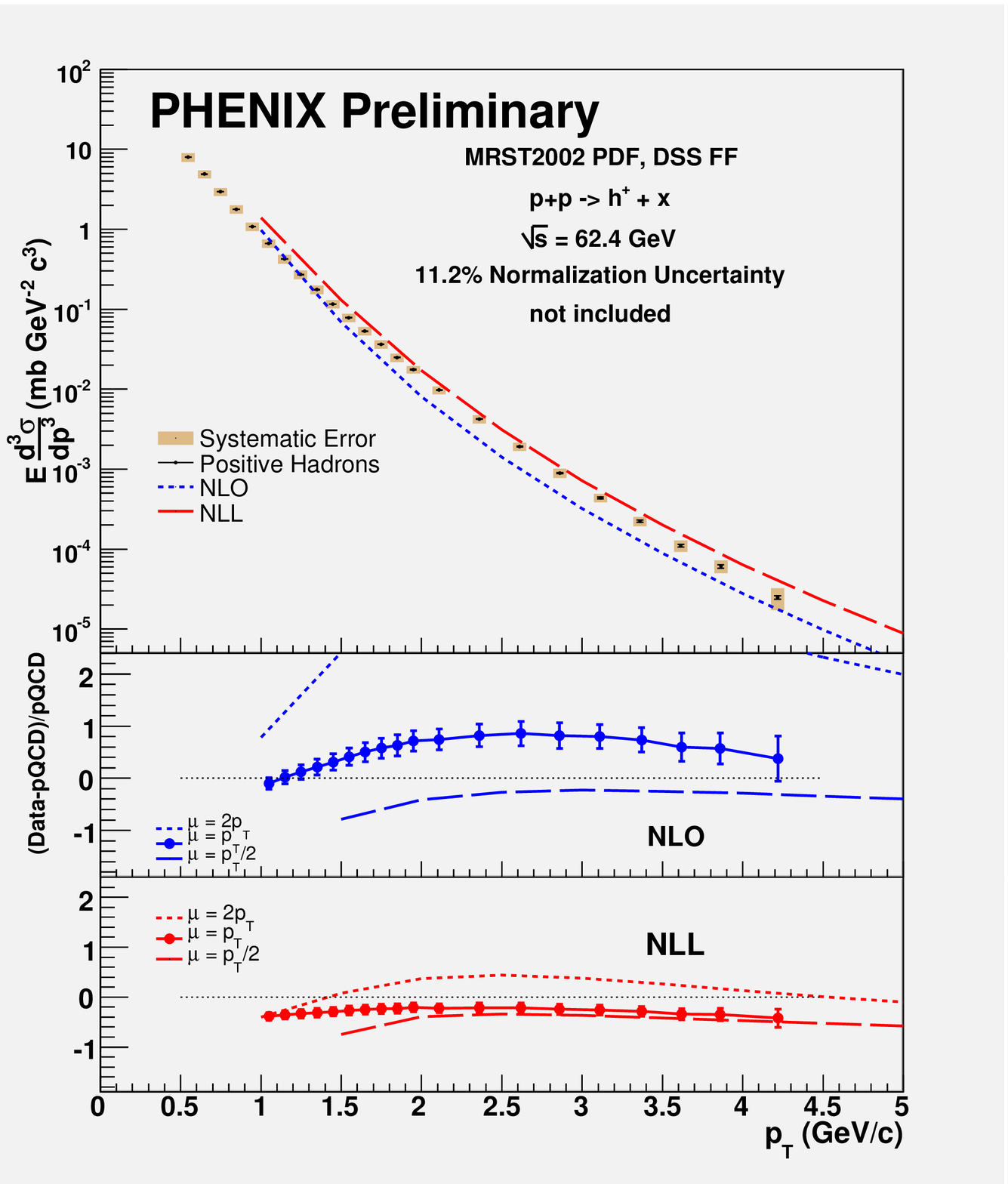}
\caption{\label{fig:crossSectionPos}Top: Cross section for positive charged hadrons produced at midrapidity in $p+p$ collisions at $\sqrt{s}=62.4$~GeV, compared to NLO and NLL pQCD calculations at theory scales of $\mu=p_T$.  Middle and bottom: The relative difference between the data and theory for three different theory scales.}
\end{minipage}\hspace{1.5pc}%
\begin{minipage}{18pc}
\includegraphics[width=18pc]{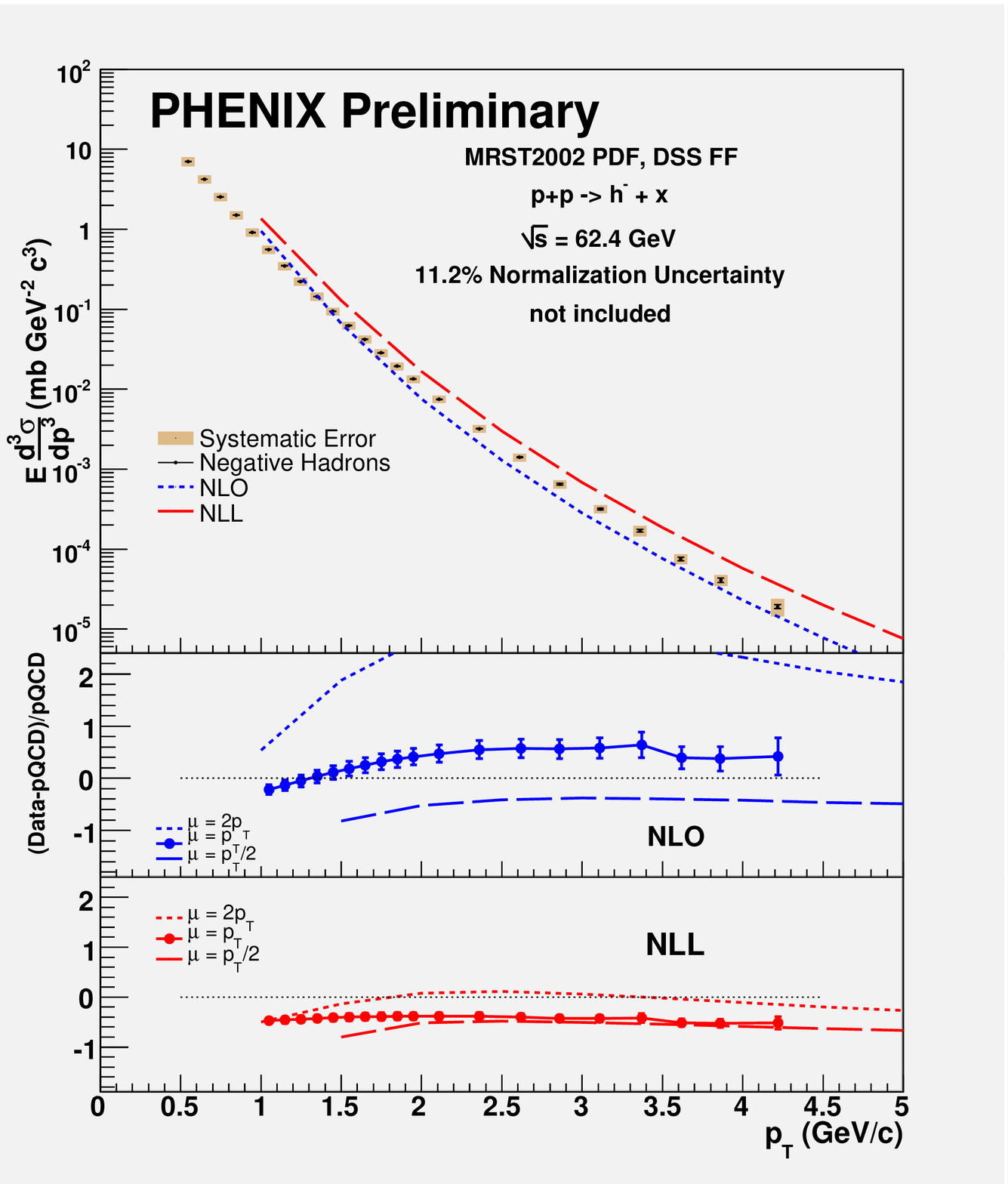}
\caption{\label{fig:crossSectionNeg}Top: Cross section for negative charged hadrons produced at midrapidity in $p+p$ collisions at $\sqrt{s}=62.4$~GeV, compared to NLO and NLL pQCD calculations at theory scales of $\mu=p_T$.  Middle and bottom: The relative difference between the data and theory for three different theory scales.}
\end{minipage}
\end{figure}

The cross section results for both positive and negative charged hadrons, shown in Figs.~\ref{fig:crossSectionPos} and \ref{fig:crossSectionNeg}, suggest that NLO pQCD calculations continue to underpredict measurements at a center-of-mass energy of 62.4~GeV, with threshold resummation offering an improved description of the data.  In particular, the dependence on the choice of factorization and renormalization scale ($\mu$) is greatly reduced in the NLL calculation.  However, it should be noted that the theoretical, as opposed to empirical, applicability of threshold resummation to midrapidity hadron production at this energy is not as clear as at fixed-target energies, as discussed in~\cite{deFlorian:2007ty}.  The DSS fragmentation functions for charged hadrons~\cite{deFlorian:2007hc} were used in the calculations.

\begin{figure}
\begin{center}
\includegraphics[width=0.5\textwidth]{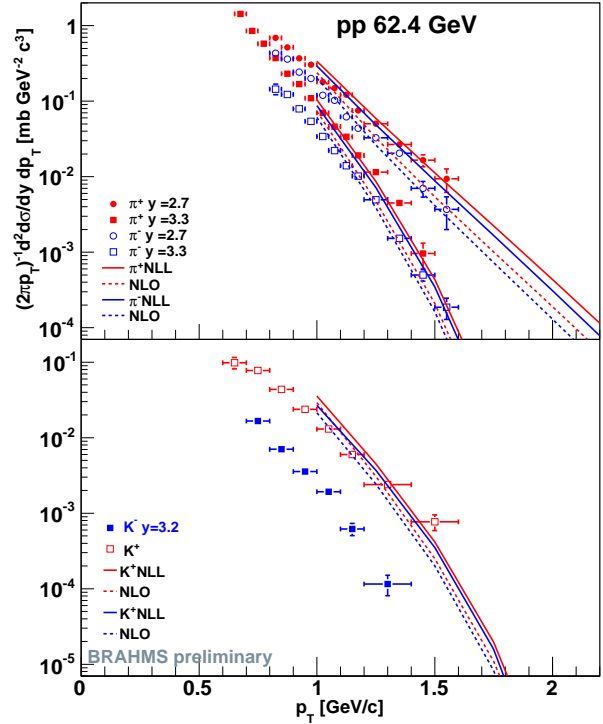}
\caption{\label{fig:Brahms}Cross sections for forward charged pion (top) and kaon (bottom) production in $\sqrt{s}=62.4$~GeV $p+p$ collisions measured by the BRAHMS experiment at RHIC compared to NLO and NLL pQCD calculations performed at a scale of $\mu=p_T$.  The data appear in \cite{Videbaek:2008zz}.}
\end{center}
\end{figure}

To emphasize the continued importance of comparing experimental data to pQCD calculations in order to improve both calculation techniques as well as input pdfs and FFs, cross section measurements for forward production of identified charged pions and kaons produced in $p+p$ collisions at $\sqrt{s}=62.4$~GeV from the BRAHMS experiment at RHIC are shown compared to NLO and NLL pQCD calculations in Fig.~\ref{fig:Brahms}.  For the pions, measurements at two different rapidities, $y=2.7$ and $y=3.3$, are shown, illustrating the strong dependence on rapidity, which is reasonably reproduced by the calculations at both NLO and NLL.  The positive kaons, measured at $y=3.2$, are also reasonably described in the measured $p_T$ range by both NLO and NLL calculations; however, the negative kaons are underpredicted by an order of magnitude, possibly indicating that the kaon FFs need to be re-examined.

\begin{figure}[h]
\begin{minipage}{18pc}
\includegraphics[width=18pc,height=0.28\textheight]{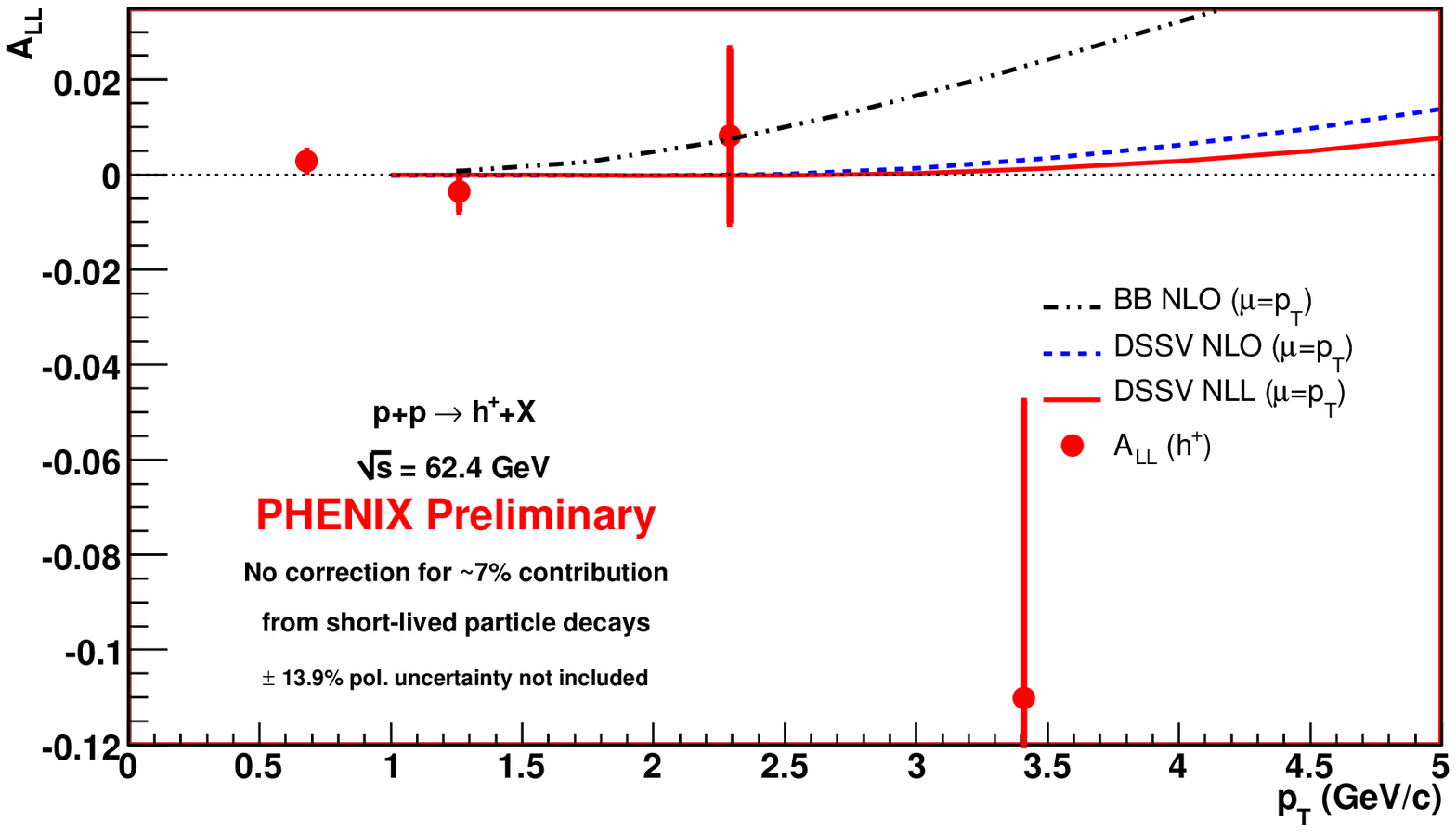}
\caption{\label{fig:asymPos}Double-helicity asymmetry for positive charged hadrons produced at midrapidity in $p+p$ collisions at $\sqrt{s}=62.4$~GeV, compared to NLO and NLL pQCD calculations at a theory scale of $\mu=p_T$.}
\end{minipage}\hspace{1.5pc}%
\begin{minipage}{18pc}
\includegraphics[width=18pc,height=0.28\textheight]{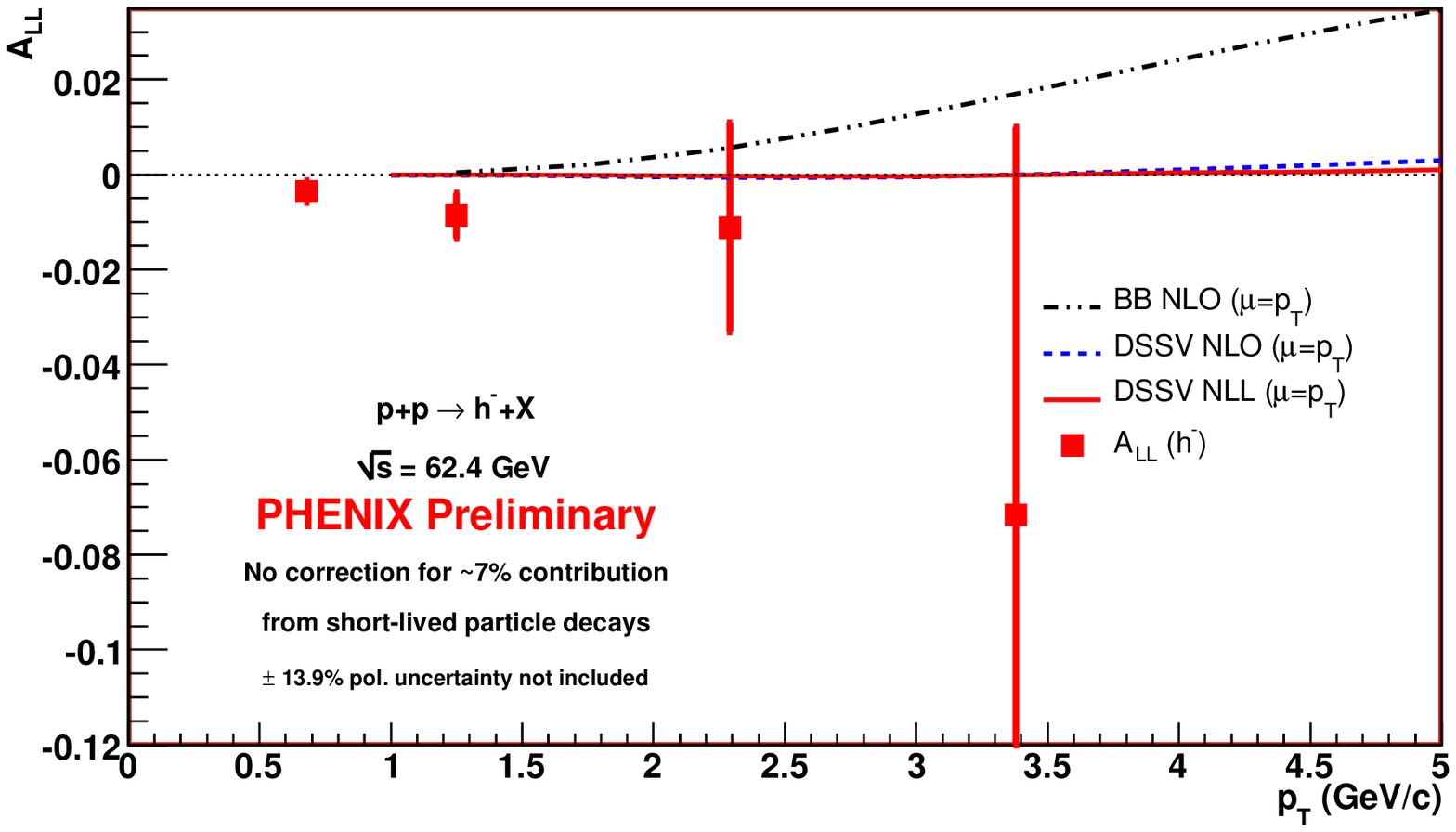}
\caption{\label{fig:asymNeg}Double-helicity asymmetry for negative charged hadrons produced at midrapidity in $p+p$ collisions at $\sqrt{s}=62.4$~GeV, compared to NLO and NLL pQCD calculations at a theory scale of $\mu=p_T$.}
\end{minipage}
\end{figure}

Figures~\ref{fig:asymPos} and \ref{fig:asymNeg} show the double-helicity asymmetry for positive and negative charged hadrons, respectively.  Similar to previous PHENIX and STAR $A_{LL}$ measurements performed at midrapidity at $\sqrt{s}=200$~GeV~\cite{Adare:2007dg,Abelev:2007vt,Adare:2008px,Abelev:2009pb,Adare:2010cy} as well as a previous PHENIX measurement of $A_{LL}$ of midrapidity neutral pions at $\sqrt{s}=62.4$~GeV~\cite{Adare:2008qb}, the results for both positive and negative hadrons are consistent with zero.

The results are compared to calculations based on polarized parton distribution functions from~\cite{Blumlein:2010rn}, denoted 'BB', and~\cite{deFlorian:2008mr}, denoted 'DSSV'.  The factorization and renormalization scales were set to the $p_T$ of the produced hadron.  The BB helicity pdf parametrization is quite recent, from 2010, but fits only inclusive deep-inelastic scattering data.  The DSSV parametrization is from 2008 and simultaneously fits inclusive and semi-inclusive deep-inelastic scattering data as well as earlier RHIC data, which suggest that the gluon helicity distribution is small in the kinematic range covered, leading to smaller predictions for $A_{LL}$.  As the cross section measurements indicate that NLL threshold resummation may improve the description of the experimental data, for DSSV, both NLO and NLL calculations are shown, with the NLL calculations being slightly smaller.  The asymmetry was calculated separately for charged pions, kaons, and protons using DSS FFs for identified particles~\cite{deFlorian:2007aj,deFlorian:2007hc}, then combined with weights reflecting the relative detection efficiency of each species as a function of $p_T$, calculated from the curves shown in Fig.~\ref{fig:eff}.  It should be noted that for the cross section measurements, this correction was done directly to the data, and non-identified charged hadron FFs were used to perform the calculations.  For the $A_{LL}$ measurement, the correction must be done to the theoretical calculations because the species dependence of the asymmetry cannot be determined from the data.

In summary, cross section and spin asymmetry measurements from RHIC over a range of energies contribute to our understanding of QCD in hadrons.

\ack
We wish to thank Daniel de Florian and Werner Vogelsang for providing calculations as well as helpful discussions.  We thank Flemming Videbaek for providing the figure from BRAHMS including the NLL calculations.

\section*{References}
\bibliography{AidalaSPIN2010Proc}

\end{document}